\documentclass[acus]{JAC2000}

\usepackage{graphicx}

\begin{document}
  \title{\flushright{WEAP038}\\[15pt]
    \centering CONTROL SYSTEM DESIGN FOR THE LIGO PRE-STABILIZED LASER}

\author{R.\,Abbott and P.\,King\\
        LIGO Laboratory\\
        California Institute of Technology, Pasadena, CA 91125, USA}

\maketitle

\begin{abstract}
To meet the strain sensitivity
requirements~\cite{abramovici},~\cite{lazzarini} of the Laser
Interferometer Gravitational Wave Observatory (LIGO), the laser frequency
and amplitude noise must initially be reduced by a factor of 1000 in the
pre-stabilized portion of the interferometer~\cite{king}.  A control system
was implemented to provide laser noise suppression, data acquisition
interfaces, diagnostics, and operator control inputs.  This paper describes
the VME-based analog and digital controls used in the LIGO Pre-stabilized
Laser (PSL).
\end{abstract}

\section{INTRODUCTION}
Gravitational waves, the ripples in the fabric of space-time, were
predicted by Einstein's General Theory of Relativity.  Although astronomical
observations have inferred the existence of gravitational waves, they have
yet to be detected directly.  The Laser Interferometer Gravitational-wave
Observatory (LIGO) is one of the large-scale gravitational-wave detectors
currently being built worldwide.

\begin{figure}[h]
  \centering
  \includegraphics*[width=\columnwidth]{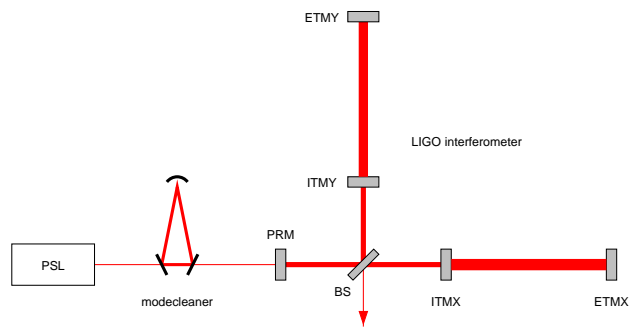}
  \caption{The LIGO interferometer.  PRM: power recycling mirror.  BS:
           beamsplitter.  ITM: input test mass.  ETM: end test mass.}
  \label{psl-position}
\end{figure}

The Pre-stabilized Laser (PSL) subsystem is the light source for the LIGO
detector as shown in Figure~\ref{psl-position}.  The output of the PSL is
modematched into the suspended modecleaner before being coupled into the
LIGO interferometer.  The term {\em pre-stabilized} is used because the
laser undergoes two stages of stabilization prior to being injected into
the interferometer.

The 10-W laser used is configured as a master-oscillator-power-amplifier
(MOPA), with a 700\,mW single-frequency, single-mode non-planar ring oscillator 
used as the master oscillator.  The control strategy uses the actuators of
the master oscillator in order to stabilize the frequency.  Power stabilization 
is achieved by control of the power amplifier output.

\section{HIGH-FREQUENCY INTENSITY NOISE SUPPRESSION}

\begin{figure}
  \centering
  \includegraphics*[width=\columnwidth]{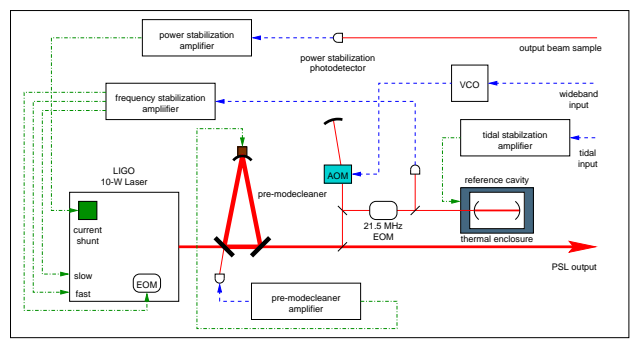}
  \caption{The components of the PSL.}
  \label{psl-parts}
\end{figure}

The PSL topology is shown in Figure~\ref{psl-parts}.  Light from the laser
is modematched into a high-throughput, ring Fabry-Perot cavity called the
pre-modecleaner (PMC).  The PSL has a design requirement that the output be
close to the shot-noise limit for 600\,mW of detected light at the
interferometer modulation frequency of 25\,MHz.  As this is beyond the
bandwidth of any electronics servo, it is done by passive filtering by the
PMC.  By appropriate choice of mirror reflectivity, the PMC acts as a tracking
bandpass filter with a pole at the cavity half-bandwidth.  One of the PMC
mirrors is epoxied to a piezoelectric transducer (PZT) to vary the length of
the cavity.  The servo electronics constantly adjusts the PZT voltage in order
to keep the incident light resonant with the cavity.

\section{FREQUENCY STABILIZATION}
Astrophysical models suggest that in order to plausibly detect candidate
gravitational-wave sources, the LIGO detector must achieve a displacement
sensitivity of better than 10$^{-19}$\,$\mbox{m}/\sqrt{\mbox{Hz}}$ at 100\,Hz.
This corresponds to a frequency noise of
10$^{-7}$\,$\mbox{Hz}/\sqrt{\mbox{Hz}}$ at 100 Hz.

\subsection{Actuators}

The frequency stabilization servo utilizes three frequency actuators inside
the 10-W laser.  A thermo-electric cooler (TEC) bonded to the laser gain
medium actuates on the laser frequency by thermally changing the optical
path length.  DC--1\,Hz adjustments to the laser frequency are made with
the TEC.  This actuator, modeled as three poles at 0.1\,Hz, has a
coefficient of 4\,GHz\,/\,V and is used for large scale adjustments to the
laser frequency.  Also bonded to the laser gain medium is a PZT, which
covers DC--10\,kHz.  A voltage applied to the PZT stresses the laser medium
and induces refractive index changes to change the laser frequency.  The PZT
has a flat response to $\sim$\,100\,kHz and is known to have a number of
mechanical resonances beyond 100\,kHz.  Fast frequency fluctuations beyond
10\,kHz are handled by the third frequency actuator, a Pockels cell located
between the master oscillator and power amplifier.

\subsection{Implementation}

A small fraction of the output of the PMC is sampled and frequency shifted
through an 80\,MHz acousto-optic modulator (AOM).  The output of the AOM is
focussed into a phase modulator that imparts sidebands at 21.5\,MHz.  The
output of the phase modulator is then modematched into a high-finesse, linear
Fabry-Perot cavity which is used as a frequency reference against which the
laser frequency is stabilized.  The frequency stabilization scheme employs the
well-known Pound-Drever-Hall technique in which the light incident on the
reference cavity is phase-modulated~\cite{drever}.  Both the carrier and
sideband light reflected from the reference cavity is focused onto a
tuned photodetector.  The output of the tuned photodetector is bandpass
filtered and synchronously demodulated to derive the error signal.

In order to ensure closed-loop stability, the open-loop gain of the PZT
actuator must be well below that of the Pockels cell at the PZT mechanical
resonance frequency.  To ensure this, the PZT actuator path is aggressively
rolled off after the designed 10\,kHz crossover.  In the absence of the
Pockels cell, the PZT path is naturally unstable at $\sim$\,15\,kHz.  With a
dynamic range some 30 times greater than that of the Pockels cell, a
self-sustaining oscillation may arise if saturation occurs in the Pockels
cell path.  Limiting the dynamic range of the PZT actuator prevents this
instability.

\section{INTENSITY STABILIZATION}
Photons in the laser light induce a source of noise in the interferometer
known as radiation pressure noise.  This noise arises from the momentum
imparted to the mirrors as statistically different numbers of photons
reflect off the mirrors in the interferometer.  To minimize the movement
of the interferometer mirrors due to radiation pressure, the intensity
fluctuations of the laser must be stabilized to the level of
$\sim$\,10$^{-8}$\,1\,/\,$\sqrt{\mbox{Hz}}$.

\subsection{Actuator}
Currently in the prototype design phase, the intensity servo utilizes a
current shunt for fast regulation of the power amplifier pump diode current.
Placed in parallel with the power amplifier pump diodes, the current shunt
was designed to carry $\sim\ \pm$\,250\,mA.

\subsection{Implementation}
\begin{figure}
  \centering
  \includegraphics*[width=\columnwidth]{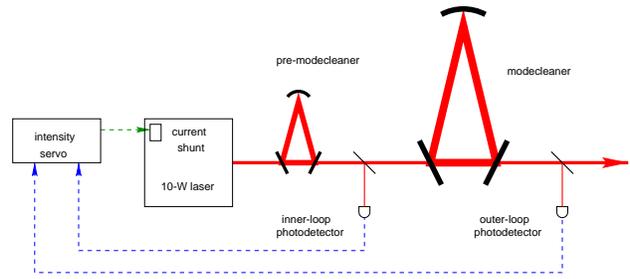}
  \caption{The intensity stabilization system layout.}
  \label{psl-iss}
\end{figure}

The intensity stabilization servo adopts a dual-loop topology as
illustrated in Figure~\ref{psl-iss}.  Inputs from photodetectors located
after the PMC and modecleaner are used in either a single or dual sensor
configuration.  In the single sensor configuration, the outer-loop
photodetector provides the signal to the servo electronics.  In the case
where the modecleaner is not locked, the single-sensor signal comes from
the inner loop photodetector.  In the dual sensor case, both the inner and
outer feedback paths provide signals to the servo electronics.

In the dual loop configuration, noise suppression is established in two
phases.  Closing the inner loop yields a high-bandwidth, well-behaved inner
loop with partial noise suppression.  The outer loop is then closed around
the inner loop to provide the balance of the noise suppression.

\section{DATA ACQUISITION AND USER CONTROL}
The user control and interface is via the Experimental Physics and
Industrial Control System (EPICS).  Through EPICS the operator can remotely
monitor the performance of the PSL and adjust the various servo loop gains
and settings.  The operator interface is a series of graphical screens, that
indicate the current status of the PSL.  Processing the data and events is
the input\,/\,output controller (IOC), a Baja4700E MIPS-based processor
running the vxWorks kernel.  The IOC performs the real-world input/output
tasks and local control tasks, and provides status information through the
Channel Access network protocol.

The control software for the PSL is event-driven and is written in state
notation language.  Although not fully debugged, automated operation from
cold start through to full operation has been demonstrated.  One software
routine constantly adjusts the TEC on the 10-W laser to keep the
laser frequency well within the dynamic range of the PZT.  One consequence
of this is that lock re-acquisition is instantaneous once the cause of the
loss of lock is removed.

At present a dozen signals are acquired and logged through the LIGO data
acquisition system.  Fast signals are acquired at the rate of 16\,kHz
whilst slower signals are acquired at 256\,Hz.  All signals are recorded
and logged.

\section{ACKNOWLEDGEMENTS}
We thank the entire LIGO team for assistance and support.  This work is
supported by the National Science Foundation under cooperative agreement
PHY--9210038.

\end{document}